\documentclass[a4paper]{iopart}
\usepackage[utf8]{inputenc}
\usepackage[dvipdfmx]{graphicx}
\usepackage{amsmath,amssymb,braket,dsfont,xcolor}
\usepackage{array}
\usepackage{float}
\usepackage[
   hmarginratio={1:1},     
   vmarginratio={1:1},     
   textwidth=457pt,        
   textheight=630pt,       
   heightrounded,          
]{geometry}

\newcommand{\be}{\begin{equation}}
\newcommand{\ee}{\end{equation}}
\newcommand{\bea}{\begin{eqnarray}}
\newcommand{\eea}{\end{eqnarray}}

\begin{document}

\title{Quantum clock models with infinite-range interactions}
\author{Adu Offei-Danso}
\address{SISSA -- International School for Advanced Studies, via Bonomea 265, 34136 Trieste, Italy.}
\address{The Abdus Salam International Center for Theoretical Physics, Strada Costiera 11, 34151 Trieste, Italy.}
\author{Federica Maria Surace}
\address{SISSA -- International School for Advanced Studies, via Bonomea 265, 34136 Trieste, Italy.}
\address{The Abdus Salam International Center for Theoretical Physics, Strada Costiera 11, 34151 Trieste, Italy.}
\author{Fernando Iemini}
\address{Instituto de F\'isica, Universidade Federal Fluminense, 24210-346 Niter\'oi, Brazil}
\address{The Abdus Salam International Center for Theoretical Physics, Strada Costiera 11, 34151 Trieste, Italy.}
\author{Angelo Russomanno}
\address{Max-Planck-Institut f\"ur Physik Komplexer Systeme, N\"othnitzer Strasse 38, D-01187, Dresden, Germany}
\author{Rosario Fazio}
\address{The Abdus Salam International Center for Theoretical Physics, Strada Costiera 11, 34151 Trieste, Italy.}
\address{Dipartimento di Fisica, Università di Napoli “Federico II”, Monte S. Angelo, I-80126 Napoli, Italy}

\begin{abstract}
We study the phase diagram, both at zero and finite temperature, in a class of $\mathbb{Z}_q$ models with infinite range interactions. We are able to identify the transitions between a symmetry-breaking and a trivial phase by using a mean-field approach and a perturbative expansion. 
We perform our analysis on a Hamiltonian with $2p$-body interactions and we find first-order transitions for any $p>1$; in the case $p=1$, the transitions are first-order for $q=3$ and second-order otherwise.

In the infinite-range case there is no trace of gapless incommensurate phase but, when the transverse field is maximally chiral, the model is in a symmetry-breaking phase for arbitrarily large fields. We analytically study the transtion in the limit of infinite $q$, where the model possesses a continuous $U(1)$ symmetry.

\end{abstract}

\section{Introduction}

Quantum clock models have recently attracted a strong interest. They display a discrete internal symmetry $\mathbb{Z}_q$ that can be spontaneously broken, in analogy with the $\mathbb{Z}_2$ symmetry of spin chains~\cite{Fradkin1980,Ostlund81}. So far, the majority of the works focused on one-dimensional short-range models~\cite{ortiz2012,zhuang2015, Sachdev_num, Sachdev_QFT, sun2019}, 
 
which are particularly interesting because of their relation with parafermionic chains: spontaneous symmetry breaking in the clock model results in non-trivial topological phases of the corresponding parafermionic model, in analogy with the relation between the Ising chain and Kitaev superconducting wire~\cite{fendley2012}. Parafermionic edge modes
are in fact the analogs of the Majorana modes of the $\mathbb{Z}_2$ symmetric models, and can be relevant for quantum computation due to their potential applicability for universal quantum computing hardware~\cite{Nayak2008}.
The implementation, however, is extremely challenging and,
also at the theoretical level, studying parafermionic chains has revealed a much more intricate problem than studying their fermionic counterparts. The first fundamental issue is that parafermionic models are intrinsically interacting, since free parafermions cannot exist in Hermitian Hamiltonians~\cite{Fendley2014}. On the other hand, their complexity offers interesting properties: Parafermionic chains can simultaneously host symmetry breaking and non-trivial topology~\cite{Bondesan2013,Alexandradinata2016}; moreover, parafermionic zero-energy edge modes can be of different nature~\cite{Jermyn2014} ("strong" or "weak" depending on whether they extend to the full spectrum or to the low-energy manifold only).

In parallel with this plethora of parafermionic phases, quantum clock models can host a wider variety of phases compared to the Ising model.
Already the simplest case with $q=3$ shows, in addition to the trivial and the symmetry-breaking phase, also a gapless incommensurate phase~\cite{zhuang2015}.
In the incommensurate phase correlations decay algebraically and are characterized by a wavelength that is incommensurate with the lattice spacing.
This rich phase diagram depends on an additional parameter, the chirality, i.e. the explicit breaking of charge conjugation symmetry~\cite{Ostlund81,Huse81,howes1983,huse1983}, which is not present in the $\mathbb Z_2$ case. Very little is known on the phase diagrams and the phase transitions of clock models with $q>3$:
for example, for $q\geq 5$ the self-dual clock models
exhibit phase transitions
of the Kosterlitz-Thouless universality class~\cite{Matsuo2006,ortiz2012,sun2019}. In general, characterizing the phase transitions of clock models has required a considerable theoretical effort and the application of advanced numerical techniques~\cite{zhuang2015, Sachdev_num, Sachdev_QFT, ChepigaMila, Giudici}.

Quantum clock models are interesting also from the point of view of experiments and applications. In a recent experiment with Rydberg atom chains~\cite{bernien2017} it has been observed that Rydberg excitations on the chains can arrange in $\mathbb{Z}_q$ ordered states, with phase transitions belonging to the same universality class as $\mathbb{Z}_q$ clocks.
Furthermore, clock models could be used for realizing exotic phases of matter, such as many-body localized phases and Floquet time crystals with arbitrary period $n$-tupling~\cite{federica,markus}: time-translation symmetry breaking can occur in disordered one-dimensional short-range clock models, but also in models with infinite-range interactions.

In this paper we inquire in more depth and generality the nature of phase transitions in the clock models with infinite-range interactions. We use a mean-field analysis which in this context is exact in the thermodynamic limit and allows us to directly study the properties of the order parameter, while numerical works in one dimension focused on other probes for the transition, like the entanglement entropy~\cite{zhuang2015}, the ground-state degeneracy or the fidelity susceptibility~\cite{sun2019}.

The model we study is a generalization of the $p$-spin model~\cite{Bapst_2012,mathfound,J_rg_2010} (with $p=1$ corresponding to the case of two-body interactions). Besides the method employed in constructing the mean-field free energy which relies on steepest descent arguments, It must be mentioned that there exists a general rigorous solution of the non-polynomial mean-field models~\cite{BRANKOV197782,Brankov1979}.
Instead of the approximation Hamiltonian method of ~\cite{BRANKOV197782,Brankov1979}, we provide
in ~\ref{sec:trotter} a calculation of the pseudo-free energy suitable for our
purposes.
We allow for an explicit breaking of charge conjugation symmetry, parameterized by the phase $\varphi$. We find that the phase structure is simpler than the one of the one-dimensional short-range model: there are a disordered phase and a broken-symmetry phase, and the transition between the two phases is either first or second order depending on $q$ and $p$. We reconstruct the phase diagram in all the cases, by finding the transition point as a function of the chirality $\varphi$.

 The paper is organized as follows. In Section~\ref{model:sec} we introduce the Hamiltonian and discuss its symmetries. In Section~\ref{free:sec} we derive the free energy density by mean-field treatment and we discuss the possible phase transitions in the light of the symmetries. 
In Section~\ref{sec:perth} we compare the numerical results concerning the continuous phase transition with the analytical results obtained via perturbation theory. We are able to derive the analytical expression of the phase-boundary line (see Fig.~\ref{phase_diagram:fig}) for $q\ge 4$. In Section~\ref{sec:piq} we discuss the fully chiral case $\varphi=\pi/q$ and we interpret the corresponding absence of the trivial phase as an exception to the analytic expansion of the free energy density introduced in Section~\ref{model:sec}. In Section~\ref{larq:sec} we consider the limit of large $q$ and study its thermodynamic properties using a harmonic approximation. We conclude and present the perspectives of future work in Section~\ref{conca:sec}. In all the paper we will assume
the Planck constant $\hbar = 1$ and the Boltzmann constant $k_B = 1$.

\section{The model} \label{model:sec}

In this section we introduce the $\mathbb{Z}_q$-invariant fully connected model (Sec.~\ref{hilbert:sec}). 
We summarize the phase structure of our model in Sec.~\ref{summary:sec}.

\subsection{Hilbert space and Hamiltonian}
\label{hilbert:sec}
\begin{figure}[h]
\begin{center}
\begin{tabular}{c} 
  \includegraphics[width=2.6cm]{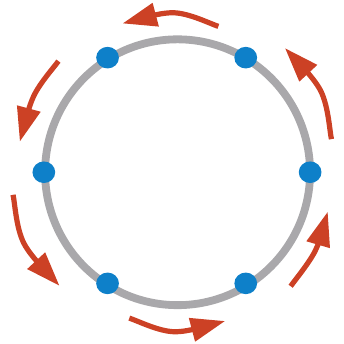}
\end{tabular}
\end{center}
\caption{Pictorial representation of a clock variable with $q=6$. It belongs to a
$q$-dimensional Hilbert space, and the blue points on the circle
indicate the possible states of the clock in the basis where $\sigma$ is diagonal. The red arrows represent
the action of the $\tau$ operators in this basis.} \label{clock:fig}
\end{figure}

Clock models generalize the Ising $\mathbb{Z}_2$ symmetry to a symmetry $\mathbb{Z}_q$ with an integer $q\ge 2$~\cite{fendley2012}. We consider a system of $N$ clock variables: each variable has $q$ possible states, that can be pictorially represented as $q$ points on a unit circle (see Fig.~\ref{clock:fig}).
We label each state with the corresponding complex number, which can assume the values $1, \omega, \omega^2,\dots, \omega^{q-1}$, where $\omega=e^{2\pi i/q}$. On the $q$-dimensional Hilbert space of a quantum clock variable we define the two operators $\hat \sigma$ and $\hat \tau$ that generalize the Pauli matrices $\hat \sigma^z$, $\hat \sigma^x$. They satisfy 

\bea
\label{zqspin1}
\hat \sigma^q = \hat \tau^q =1\ , \qquad \hat \sigma^\dagger &=& \hat \sigma^{q-1}\ , \qquad \hat \tau^\dagger = \hat \tau^{q-1}\ ,\\
\hat \sigma \hat\tau = \omega\, \hat \tau \hat \sigma\ . & & 
\label{zqspin2}
\eea
A convenient representation for the operators is the following

\be \hat \sigma = 
\begin{pmatrix}
1&0&0&\ \dots\  & 0\\
0&\omega&0&\ \dots\  & 0\\
0&0&\omega^2&\   & 0\\
\vdots&\vdots&\vdots&&\vdots \\
0&0&0&\ \dots\  & \omega^{q-1}
\end{pmatrix},\quad\quad
 \hat \tau = 
\begin{pmatrix}
0&0&0& \dots\  &0& 1\\
1&0&0& \dots\  &0 & 0\\
0&1&0& \dots   & 0 &0\\
\vdots&\vdots&\vdots&&\vdots&\vdots \\
0&0&0& \dots\ & 1& 0
\end{pmatrix}
\label{explicitrep}
\ee
In this representation, $\hat \sigma$ measures the position on the unit circle, and $\hat\tau$ shifts the state of one position counter-clockwise along the circle. In the case $q=2$, the matrices in Eq.~(\ref{explicitrep}) coincide with the canonical Pauli matrices $\hat \sigma_z$, $\hat \sigma_x$.

We define a Hamiltonian for $N$ sites, in terms of $\hat \sigma_j$ and $\hat \tau_j$ acting on site $j$. On the same site the operators satisfy the relations in Eqs.~(\ref{zqspin1},\ref{zqspin2}), and on different sites they commute.
We define the two operators
\be
\hat{m}_{\sigma} =\frac{1}{N} \sum_{j=1}^N\hat\sigma_{j}\qquad
\hat{m}_{\tau} = \frac{1}{N} \sum_{j=1}^N\hat\tau_{j}
\ee
which represent the total ``magnetizations'' along $\hat \sigma$ and $\hat \tau$. The Hamiltonian of our fully connected model is then defined as
\be\label{fcham}
\hat H = -N\left(\hat{m}_{\sigma}\hat{m}_{\sigma}^{\dagger} \right)^{p} - h q^2 N\left(\hat{m}_{\tau}e^{i\varphi} + \hat{m}_{\tau}^{\dagger}e^{-i\varphi}\right),
\ee
where $p \ge 1$, $h\ge 0$ is the transverse field, $\varphi$ is real and the factors $N$ guarantee the extensivity.\newline The case $q=2$ and $\varphi = 0 $ corresponds to the fully connected $p$-spin ferromagnet~\cite{Bapst_2012}. Before proceeding, we briefly outline the main features for this case. In the limit of large $h$, one observes a paramagnetic $\mathbb{Z}_2$ invariant state. For $h$ below a critical value, on the opposite, the system chooses between two broken-symmetry states which physically correspond to a ferromagnet with all spins pointing either up or down in the $z$ direction. The nature of the transition separating these two phases is second order for $ p = 1 $ and first order for $p >1$. The case of $ p \rightarrow \infty $ is connected to Grover's search algorithm~\cite{grover1997quantum}. \newline Qualitatively one would expect a similar behaviour in the case of $ q > 2 $ and $\varphi = 0$ (i.e. $q$ broken symmetry states for $ h \rightarrow 0 $, and a  $\mathbb{Z}_q$  invariant para-magnetic phase for $ h \rightarrow \infty$). The crux of this paper will be the elucidation of the nature of the phase transitions for $q \ge 2$  and the interesting behaviour that arises with the introduction of chirality ($\varphi \ne 0$). To this end, it is important to discuss the symmetries of the model.

The Hamiltonian (\ref{fcham}) has a global $\mathbb{Z}_q$ symmetry generated by the unitary operator
\be\label{symm}
\hat G=\prod_{j=1}^N \hat\tau_j.
\ee
We can also notice that the Hamiltonian is invariant under time reversal, which is defined as the antiunitary transformation
\be
\hat T \hat \sigma_j \hat T = \hat \sigma_j^\dagger,\qquad \hat T \hat \tau_j \hat T =\hat \tau_j,\qquad \hat T^2=\mathbb{I}. 
\ee
We introduce the charge conjugation unitary operator
\be
\hat C \hat \sigma_j \hat C=\hat \sigma_j,\qquad \hat C \hat \tau_j\hat C=\hat \tau_j^\dagger, \qquad \hat C^2=\mathbb{I}.
\ee
This transformation is a symmetry only for $\varphi=0$. We refer to this special case as the non-chiral clock model, while the parameter $\varphi$ is called chirality. In general, charge conjugation transforms the Hamiltonian by changing sign to the chirality: $\hat C \hat H(\varphi) \hat C =\hat H(-\varphi)$. 
The global operator
\be
\hat K=\prod_{j=1}^N \hat\sigma_j^\dagger.
\ee
transforms the Hamiltonian as $\hat K^{-1} \hat H(\varphi) \hat K = \hat H(\varphi+2\pi/q)$. Therefore, using the combined action of $\hat C$ and $\hat K$, we can restrict without loss of generality to the case $0\le \varphi\le \pi/q$.

\subsection{Summary of the results}
\label{summary:sec}
We find that the phase diagram of the model in Eq.~(\ref{fcham}) contains a trivial phase and a symmetry-breaking phase.

\begin{figure}[h]
\begin{center}
  \includegraphics[width=12cm]{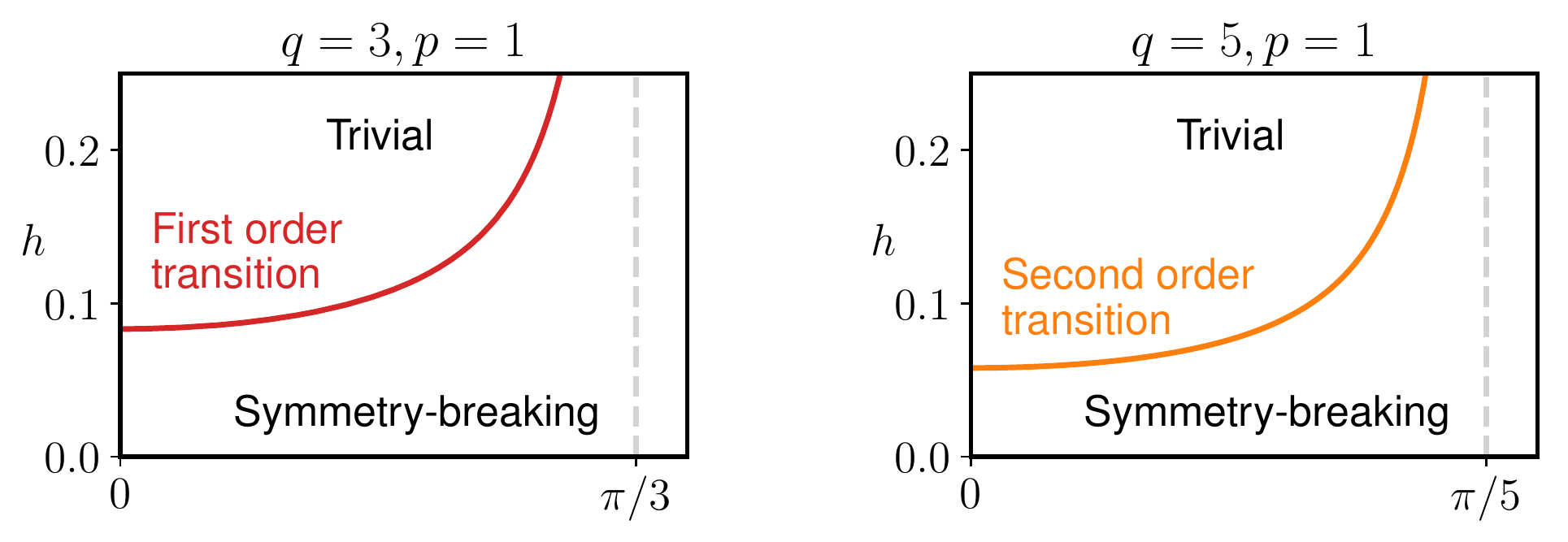}
\end{center}
\caption{(Left panel) Phase diagram for the model with $q=3$ and $p=1$. Notice the first-order transition between symmetry-breaking and trivial phase and the phase-boundary tending to $h\to\infty$ for $\varphi\to\pi/3$. (Right panel) Phase diagram for $q=5$ and $p=1$. Now the transition is second order (a fact true for all $q>3$) and at $\varphi=\pi/5$ the transition moves to infinity. For generic $q$ this fact occurs at the fully chiral point $\varphi=\pi/q$.} \label{phase_diagram:fig}
\end{figure}

For $q=3$, $p=1$ the transition between the trivial phase and the symmetry-breaking phase is first order (we show the phase diagram in Fig.~\ref{phase_diagram:fig}). 
The most peculiar point is at chirality $\varphi=\pi/3$: The value of the field at the transition goes to infinity as we approach $\varphi=\pi/3$, and the system is always in a broken symmetry phase for that value of $\varphi$.  

For any $q>3$, $p=1$ in the infinite-range model there is a second-order transition from symmetry-breaking to trivial phase. We show the phase diagram for $q=5$ in Fig.~\ref{phase_diagram:fig}. 
The phase-boundary curve is given by the analytical formula in Eq.~\eqref{eq:hc}.

For $p>1$, on the other hand, the transition is of first order for any value of $q$. In all these cases, when $\varphi=\pi/q$ (fully chiral case) and the temperature is below a threshold ($T\le 1/2$)  we still see that only the symmetry-breaking phase exists in our model. 
Remarkably, this result shows that the chirality and the explicit breaking of the charge conjugation symmetry have a deep influence on the thermodynamic properties also in this infinite-range interacting context.

\begin{center}
\begin{tabular}{ | m{5cm} | m{3cm}| m{3cm} | } 
\hline
 $(q,\varphi)$  & $p = 1$  & $p > 1$ \\ 
\hline

$q = 3,\varphi \neq \frac{\pi}{q} $ &1st order & 1st order \\ 
\hline
$q = 3,\varphi = \frac{\pi}{q} $ & No transition & No transition \\ 
\hline
$q > 3,\varphi \neq \frac{\pi}{q}$ & 2nd order & 1st order \\ 
\hline
$q > 3,\varphi = \frac{\pi}{q}$ & No transition & No transition \\ 
\hline
\end{tabular}
\end{center}

\section{Free energy}
\label{free:sec}

In this section, we study the free-energy density $f(\beta,h)$ of the model at inverse temperature $\beta$ in the thermodynamic limit, for a generic $q\ge 2$. 

Thanks to the full connectivity of the interactions, a mean-field analysis provides a good description for the statistical-mechanical properties of the system. The canonical prescription for the mean-field approach on a quantum model involves first transforming the quantum partition function $Z$ into a classical one by means of Suzuki-Trotter decomposition~\cite{Suzuki1976}. We introduce the order parameter, defined as $m =(|m|,\theta)=\braket{\hat m_\sigma}$, for the mean-field analysis and we apply the static approximation in order to remove the time dependence of the order parameter (see~\ref{sec:trotter} for details). The free energy density of our model as calculated by this procedure is given by
\be
\label{eq:f}
 f=(2p-1)|m|^{2p}+f_{s}.
\ee
with
\begin{eqnarray}
f_s & = &-\frac{1}{\beta}\log \text{Tr } e^{-\beta \hat H_{s}} \\
 \hat H_{s} & = & -(\lambda^*\hat\sigma +\lambda\hat\sigma^\dagger)-hq^2(\hat\tau e^{i\varphi}+\hat\tau^\dagger e^{-i\varphi}). \label{eq:mf}
\end{eqnarray}
where $\hat H_{s}$ corresponds to a single-site Hamiltonian, and
 the complex number $\lambda=  p m|m|^{2p-2}$ is an effective longitudinal field that depends on the average magnetization $m=\braket{\hat m_\sigma}$.

Computing the function $f_{s}$ requires the diagonalization of a $q\times q$ Hermitian matrix. However, building on the Landau theory of phase transitions, general considerations can be formulated based on the symmetries of the model. As will become clear, we further need the assumption that $f_{s}$ is an analytic function of $\lambda$ and $\lambda^*$ close to the point $\lambda=\lambda^*=0$. In the following subsections we qualitatively discuss the expansion of the free energy density $f_{s}$ as a power series in $\lambda$ and $\lambda^*$, and we examine the case where the assumption of analyticity is not valid. In both cases, these arguments are sufficient to determine if a phase transition occurs and whether it can be continuous. Quantitative results concerning the expansion of the free energy density will be obtained using perturbation theory and are discussed in Section \ref{sec:perth}.

\subsection{Series expansion}\label{sec:exp}

The single-site Hamiltonian Eq.~\eqref{eq:mf} transforms under the unitary operator $\hat \tau$ and under time reversal $\hat T$ as
\be
\hat \tau \hat H_{s}(\lambda,\lambda^*) \hat \tau^\dagger=\hat H_{s}(\omega \lambda,\omega^*\lambda^*),\qquad \hat T \hat H_{s}(\lambda,\lambda^*) \hat T=\hat H_{s}(\lambda^*,\lambda).
\ee
Since these transformations leave the trace of $\exp(-\beta \hat H_{s})$ invariant, the free energy density $f_{s}$ has to satisfy the following properties
\be
f_{s}(\lambda,\lambda^*)=f_{s}(\omega \lambda,\omega^*\lambda^*),\qquad f_{s}(\lambda,\lambda^*)=f_{s}(\lambda^*,\lambda).
\ee
As a consequence, the only non-zero terms that can appear in the power series are of the form $[\lambda^q+(\lambda^*)^q]^j(\lambda\lambda^*)^k$, for generic integers $j,k$. To lowest power in $|m|$ the free energy density $f_{s}$ reads
\be
f_{s}\simeq a_0+a_2\lambda\lambda^*=a_0+  a_2p^2|m|^{4p-2}
\ee

\begin{figure}
\centering
\includegraphics[width=0.32\textwidth]{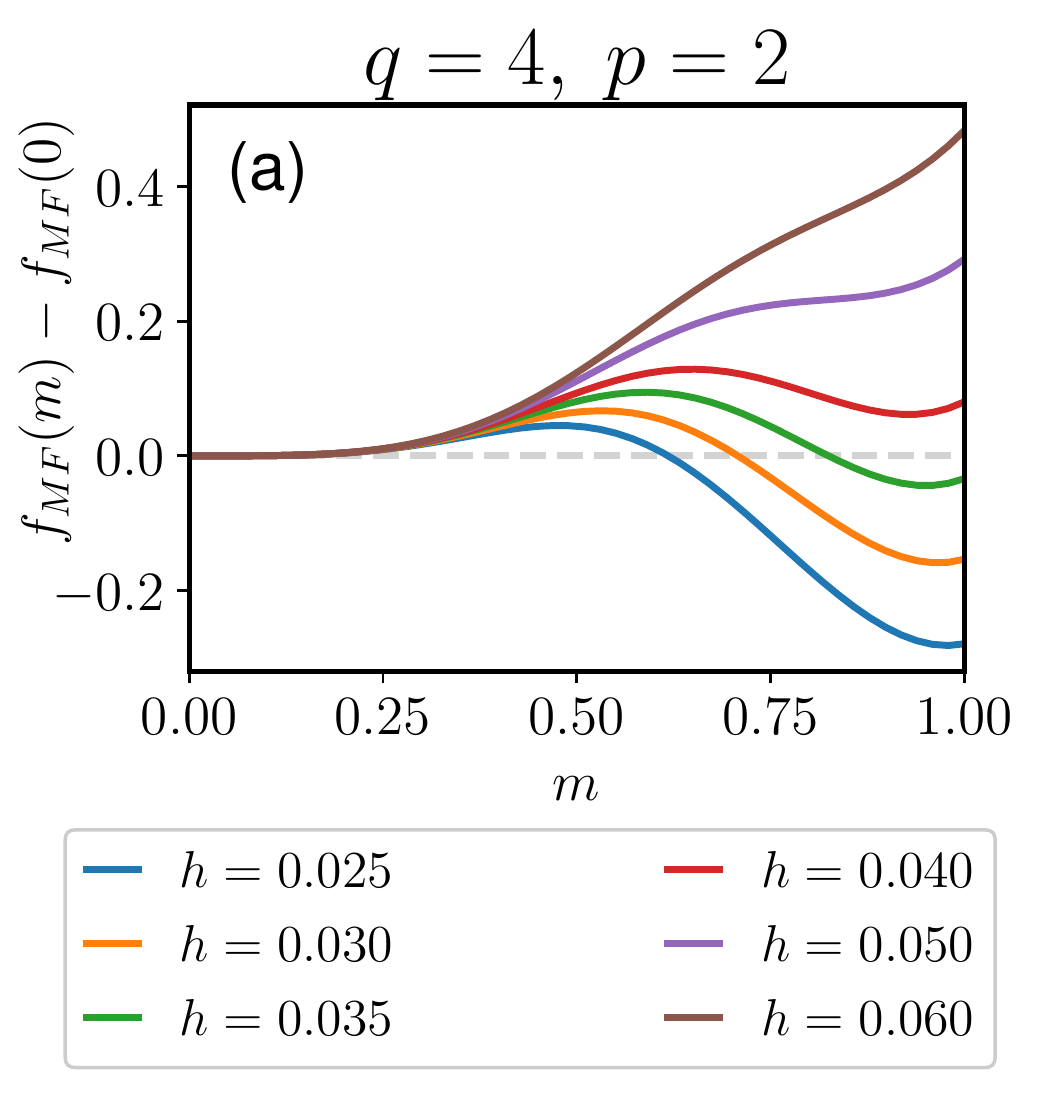}
\includegraphics[width=0.32\textwidth]{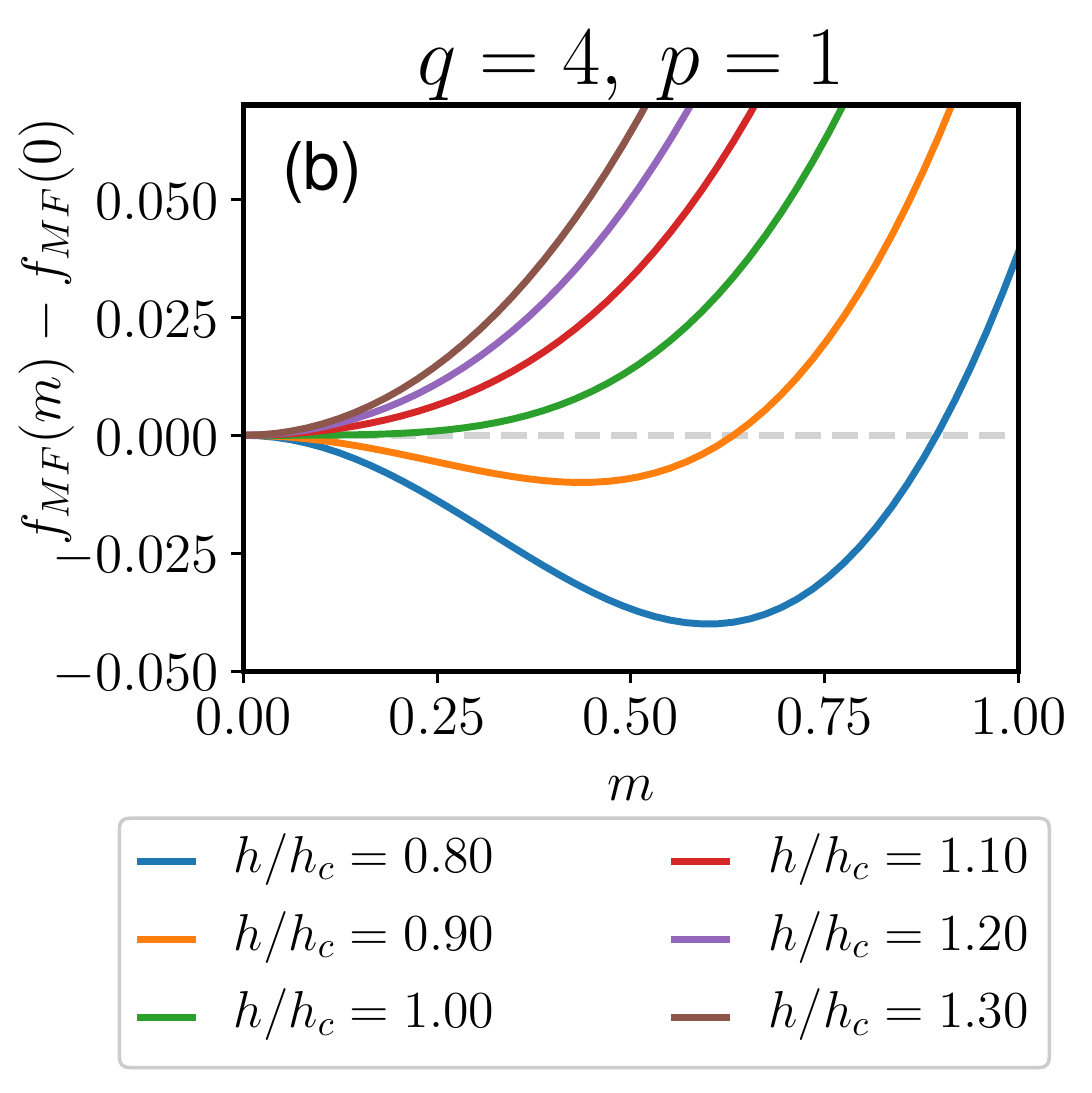}
\includegraphics[width=0.32\textwidth]{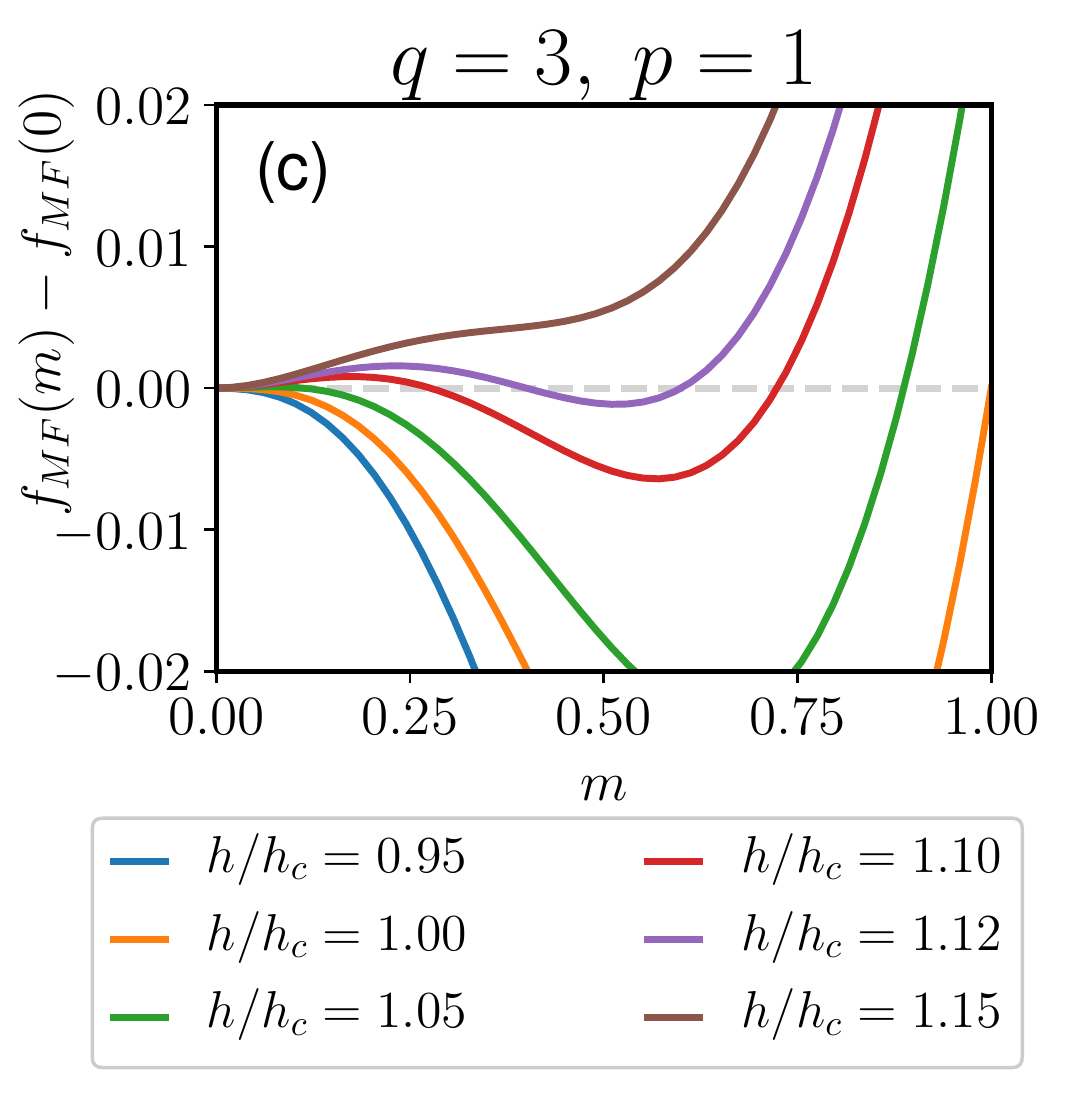}
\caption{ Free energy density at zero temperature as a function of $m$ for $\varphi=0$. (a) For $p>1$ the transition from an ordered phase to a disordered one is of first order. (b) For $p=1$, $q\ge 4$ the transition is of second order. (c) For $p=1$, $q=3$ the transition is of first order. The value of the field where the concavity in $m=0$ changes sign is called $h_c$.}\label{fig:f}
\end{figure}

Using this relation in Eq.~\eqref{eq:f}, we see that:
\begin{itemize}
    \item For $p>1$, the most relevant term in the limit $|m|\rightarrow 0$ is $(2p-1)|m|^{2p}$, which is always positive. This means that $m=0$ is a local minimum, and the phase transition to an ordered phase with $|m|\neq 0$ can only be of first order (see Fig~\ref{fig:f}-a).
    \item For $p=1$, on the other hand, the dominant term is $(1+ a_2)|m|^2$, so a continuous phase transition is in principle possible when $a_2=-1$ (see Fig.~\ref{fig:f}-b). Another possibility is to have, as we vary $h$, a regime where $a_2>-1$ (so $m=0$ is locally a minimum) but a lower global minimum appears for $m\neq 0$ (see Fig.~\ref{fig:f}-c). 
    We postpone to Section \ref{sec:perth} a more detailed discussion about the order of the phase transition in this case.
\end{itemize}

\subsection{Non-analytic behaviour}\label{sec:nonan}
The results of the previous section crucially depend on the assumption that $f_{s}$ is an analytic function of $\lambda$ and $\lambda^*$ close to the point $|\lambda|=0$. We now show a case where this assumption is not valid (due to the chirality $\varphi$) and discuss the consequences on the properties of the phase transition.

Let us consider the case of $q>2$ and zero temperature, for which $f_{s}$ is equal to the ground state energy of $\hat H_{s}$, and examine how this energy depends on the small fields $\lambda, \lambda^*$. For $|\lambda|=0$, $H_0\equiv\hat H_{s}(\lambda=0, \lambda^*=0)$ is diagonal in the $\tau$ basis and has eigenvalues $-2hq^2\cos(2\pi j/q+\varphi)$ for $j=0,1,\dots, q-1$. If the ground state of $H_0$ is unique (i.e. for $\varphi\neq \pi/q$), the first perturbative correction to the ground state energy is of second order (proportional to $\lambda\lambda^*$). On the other hand, if $\varphi= \pi/q$, the ground state of $H_0$ has double degeneracy. The Hamiltonian $H_{s}$ restricted to the ground state manifold has the form
\begin{equation}
    H_{s}|_{GS}=\begin{pmatrix}
    \epsilon_0 & -\lambda^*\\
    -\lambda & \epsilon_0
    \end{pmatrix}
\end{equation}
with $\epsilon_0=-2hq^2\cos(\pi /q)$. We obtain that, to lowest order in $|\lambda|$, the ground state energy is $f_{s}\simeq \epsilon_0 -|\lambda|$, which is not an analytic function of $\lambda$ and $\lambda^*$.

We deduce that, while the discussion of the previous section applies almost everywhere, a different scenario appears at zero temperature for $\varphi=\pi/q$. In this case, the free energy density in Eq.~\ref{eq:f} reads
\begin{equation}
f=(2p-1)|m|^{2p}-2hq^2\cos(\pi /q)-p|m|^{2p-1}+O(|m|^{4p-2}/2hq^2).
\end{equation}
Note that, in the limit of large $h$, we can neglect higher order terms, and the minimum is found for $|m|=1/2$. Remarkably, the model does not have a transition to a paramagnet, and the magnetization remains finite for arbitrarily large field $h$. This peculiar behaviour is  further discussed in Section~\ref{sec:piq}.

\section{Continuous phase transition}\label{sec:perth}

\begin{figure}
\centering
\includegraphics[width=0.30\textwidth]{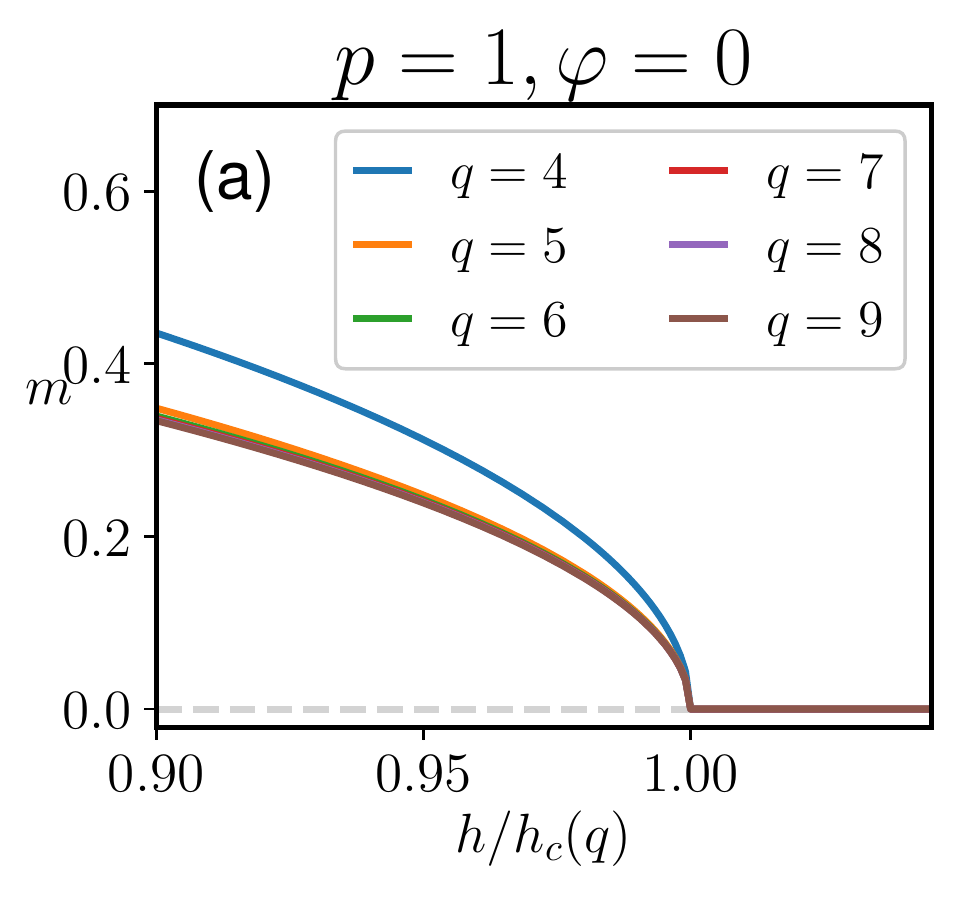}
\includegraphics[width=0.30\textwidth]{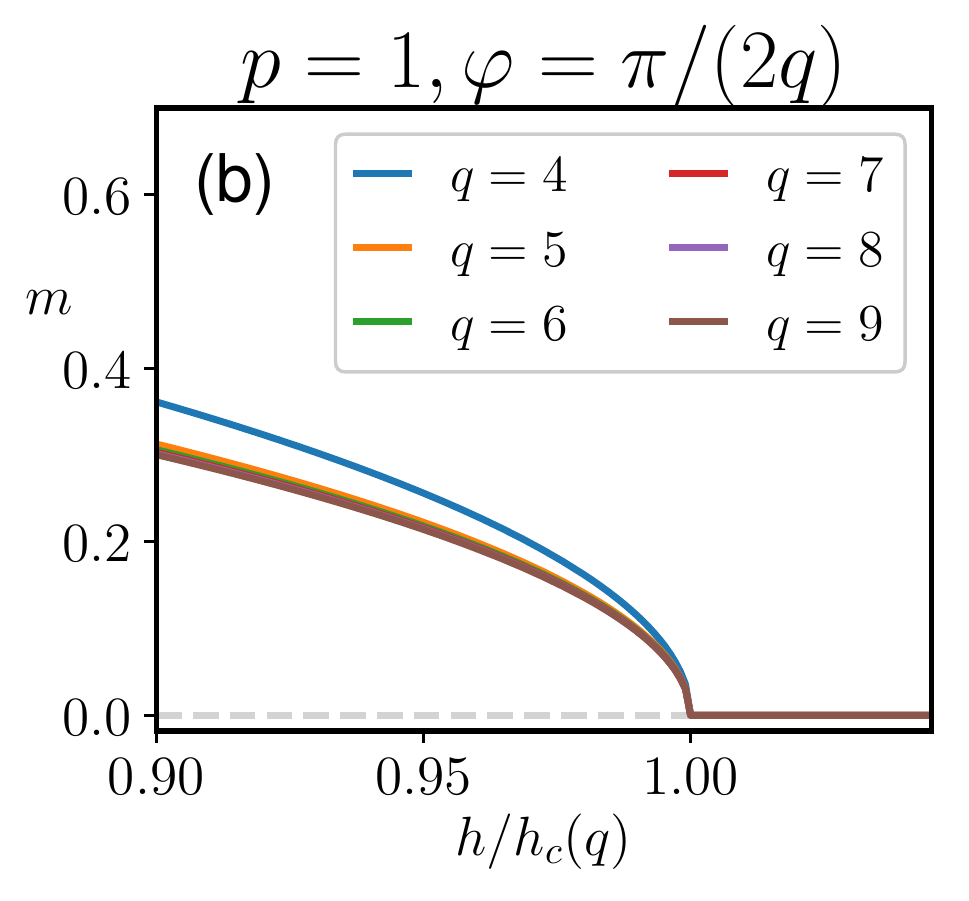}
\includegraphics[width=0.38\textwidth]{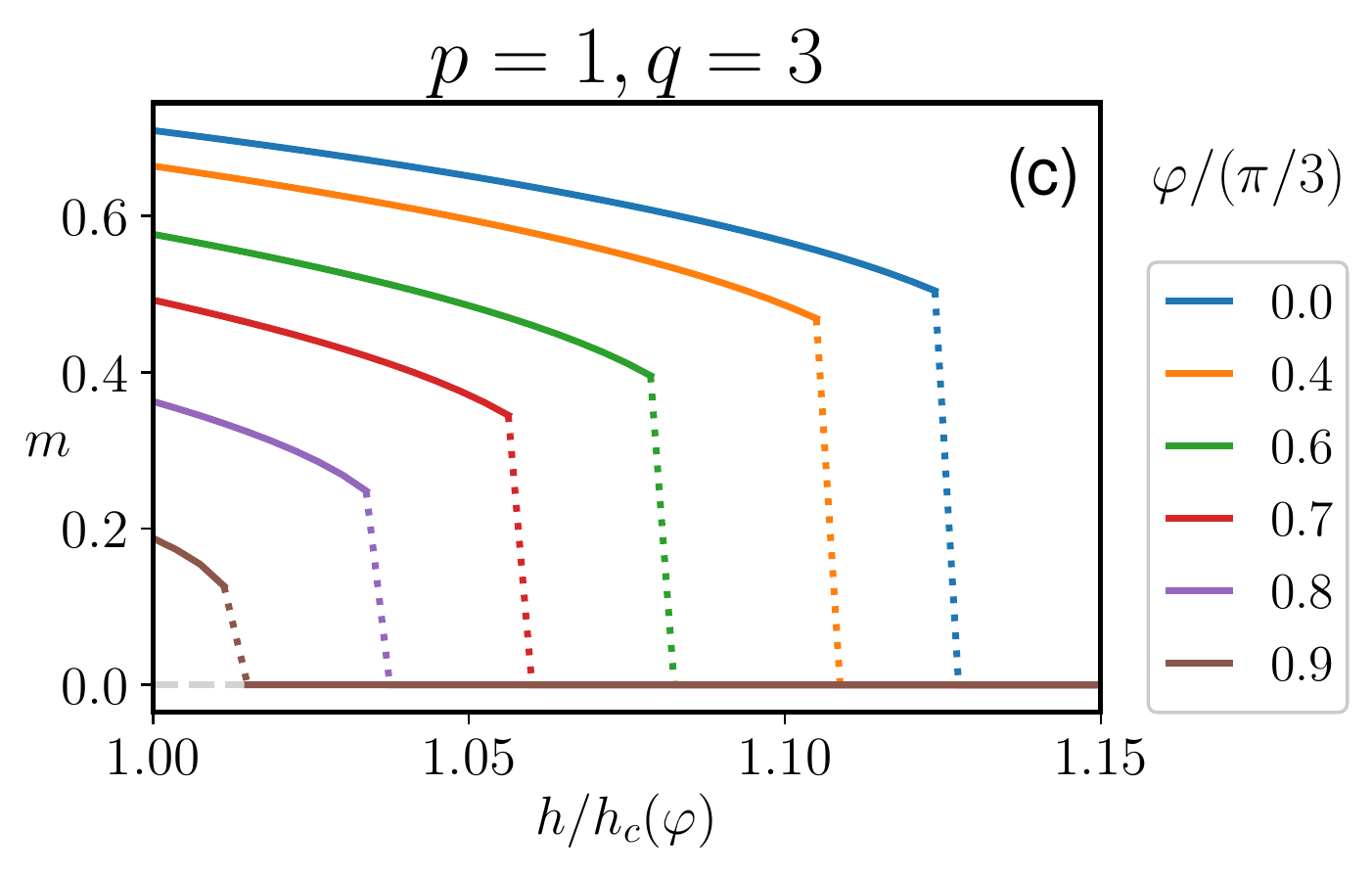}
\caption{Magnetization $m$ as a function of $h/h_c$ (from Eq.~\ref{eq:hc}). (a),(b)  A second order phase transition occurs at $T=0$, $h=h_c$ for different values of $q>3$ and $\varphi<\pi/q$. (c) A first order phase transition takes place at $T=0$, $h=h_*>h_c$ for $q=3$ and different values of $\varphi<\pi/q$.}\label{fig:mp1}
\end{figure}

Since a continuous phase transition has already been ruled out for $p>1$, we will focus from now on on the case $p=1$. As explained in the previous section, if a continuous phase transition occurs we can obtain the exact location in the phase diagram from the condition $a_2=-1$. The coefficient $a_2$ can be computed exactly using perturbation theory, for arbitrary field $h$ and inverse temperature $\beta$ (explicit calculations are reported in~\ref{app:pertth}). In particular, as we show in Fig.~\ref{fig:mp1}-a and \ref{fig:mp1}-b, the zero temperature transition is located at
\begin{equation}\label{eq:hc}
    h_c=\frac{1}{2q^2}\left[\left(\cos(\varphi)-\cos\left(\varphi+\frac{2\pi}{q}\right)\right)^{-1}+\left(\cos(\varphi)-\cos\left(\varphi-\frac{2\pi}{q}\right)\right)^{-1}\right],
\end{equation}
while the transition at zero field occurs at $\beta_c=1$.

There is, however, another possibility: the transition may be a discontinuous first-order one and may occur at a value of the field $h_*>h_c$ (or $\beta_*>\beta_c$). We argue that this is indeed the case for $q=3$. In this case, the free energy density has a third order term $\propto 2|\lambda|^3\cos(3\theta)$ which is negative for some values of $\theta=\arg(\lambda)$, and a fourth order term, which is always positive.
Given these signs of the coefficients, it can be proven 
that for $h\rightarrow h_c^+$ the difference of the free energy densities $f_{MF}(m)-f_{MF}(0)$ becomes negative for certain values of $m$ (see \ref{app:firstorder}). Therefore, $m=0$ is not the global minimum: a first order phase transition occurs for a value $h_*>h_c$ (which we obtain numerically) at $T=0$, as shown in Fig.~\ref{fig:f}-c and Fig.~\ref{fig:mp1}-c. For any other value of $q$, the third order coefficient is zero, and we expect the transition to be continuous (Fig.~\ref{fig:mp1}-a,b).

\section{Case $\varphi=\pi/q$}\label{sec:piq}
\begin{figure}
\centering
\includegraphics[width=0.4\textwidth]{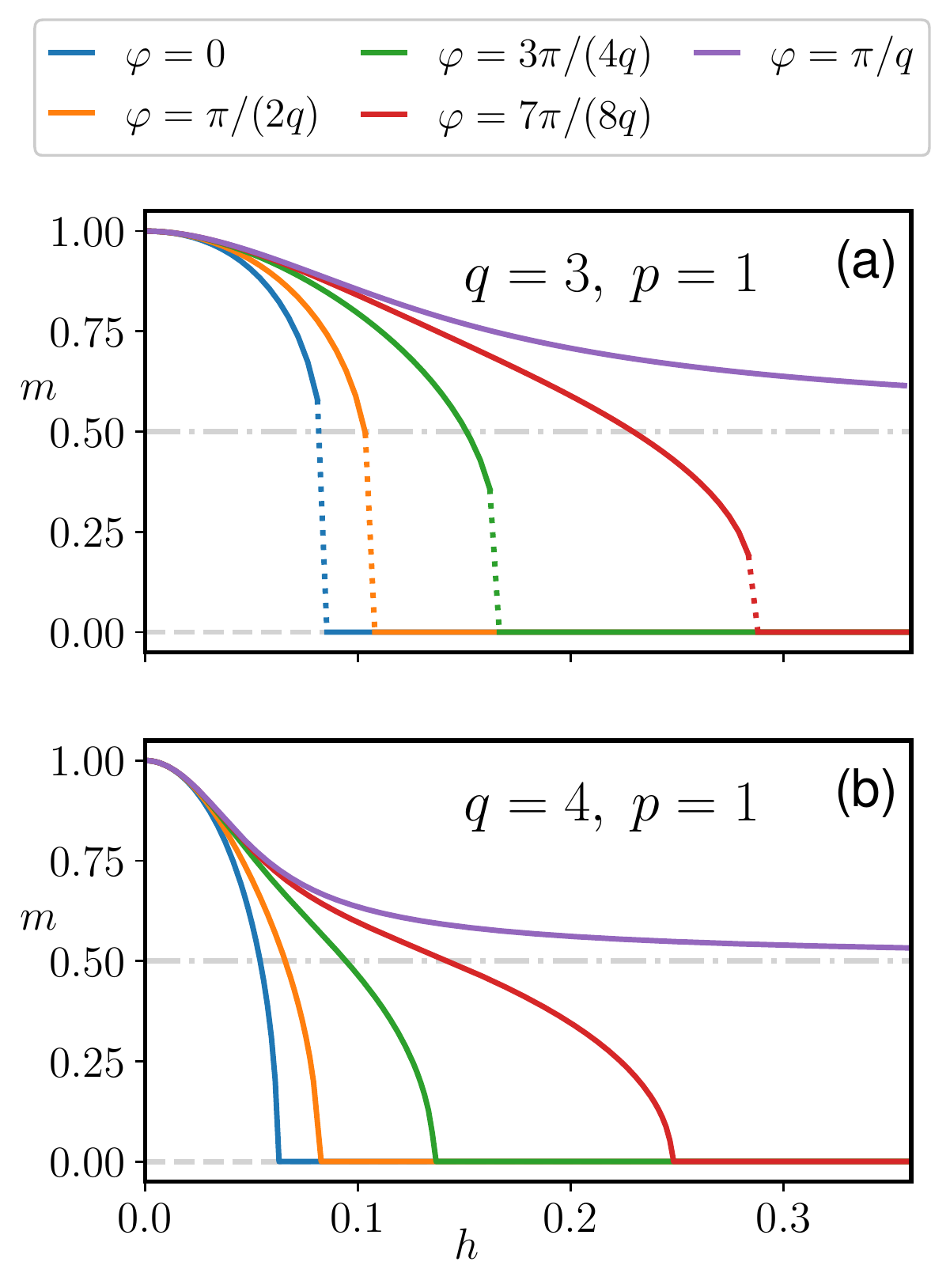}\hspace{0.5cm}
\includegraphics[width=0.4\textwidth]{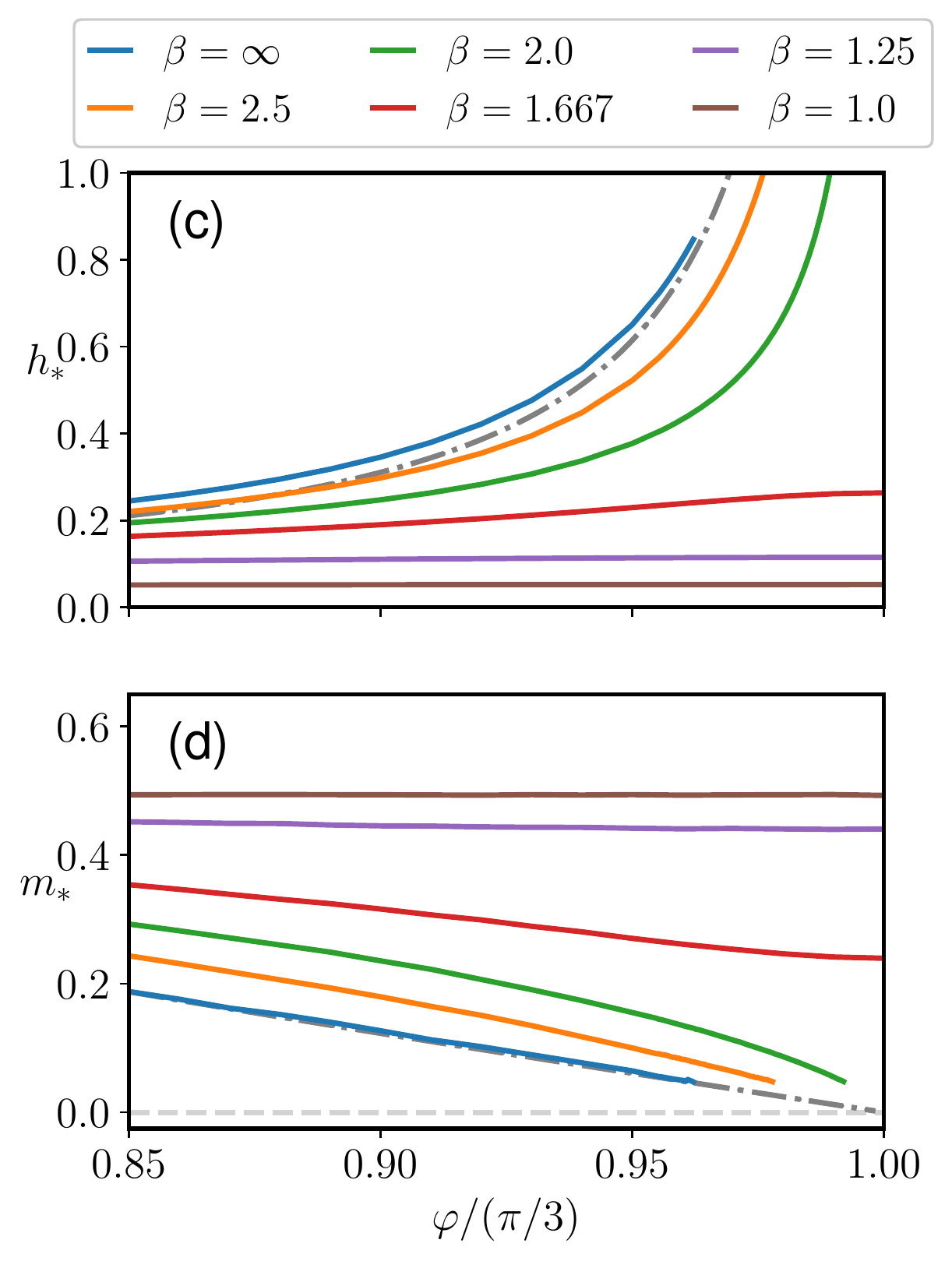}
\caption{Magnetization $m$ as a function of the field $h$ for different values of $\varphi$ in the cases (a) $q=3$, $p=1$ (first order transition) and (b) $q=4$, $p=1$ (second order transition). Field $h_*$ (c) and magnetization $m_*$ (d) at the discontinuous transition for $q=3$ as a function of the chirality $\varphi$ for different values of $\beta$.}\label{fig:pi3}
\end{figure}

From Eq.~\eqref{eq:hc} we see that the zero-temperature critical field diverges when $\varphi\rightarrow \pi/q$. We have further proved in section \ref{sec:nonan} that no transition occurs for $\varphi=\pi/q$, in which case the magnetization tends to $m\rightarrow 1/2$ for $h \rightarrow \infty$. We illustrate this non-analytic behaviour in Fig.~\ref{fig:pi3}-a: both for a discontinuous and for a continuous transition, as we approach the value $\varphi=\pi/q$, the fields at the transition ($h_*$ and $h_c$ respectively) diverge. Moreover, for the discontinuous case, the jump of the magnetization at the transition ($m_*$) tends to zero. The asymptotic behaviours at $T=0$ for $x=\pi/q-\varphi\ll 1$ read
\begin{equation}
    h_c \simeq \frac{1}{4q^2\sin(\pi/q)x}
\end{equation}
and for $q=3$
\begin{equation}
    h_* \simeq h_c\left(1+\frac{4}{3}x^2\right)
\hspace{2cm}
    m_*=36h_* x^2.
\end{equation}

 As can be seen in Fig.~\ref{fig:pi3}-a, for $\varphi=\pi/q$, the magnetization is always larger than $1/2$.

We now consider the case of $T\neq 0$ but small compared to $h$, such that $h\cdot x \ll \beta^{-1} \ll h$. In this case, the perturbative expansion can be used and
\begin{equation}
    a_2\simeq -\frac{\tanh{\left(2\beta h q^2 \sin(\pi/q)x\right)}}{4hq^2 \sin(\pi/q)x} \simeq -\beta/2
\end{equation}
so if $\beta \ge 2$ for $p=1$ the system is ferromagnetic in this regime. The phase transition can only occur out of this range, i.e. at a value of $h$ diverging at least as fast as $1/x$. Since the transition point moves to $h\rightarrow \infty$ as $x\rightarrow 0$, we can argue that for $\beta\ge 2$, as already discussed in the zero-temperature case, when $\varphi=\pi/q$ ($x=0$) the system is always ferromagnetic. This is in fact shown in Fig.\ref{fig:pi3}-b for $q=3$, where a qualitative difference can be observed between $\beta\ge 2$ and $\beta < 2$. For $\beta \ge 2$ the transition point moves to diverging values of the field $h_*$ when $x\rightarrow 0$, but it tends to a finite value when $\beta < 2$.

\section{Large $q$ limit}\label{larq:sec}
In this section, we derive  an analytic expression for the free energy density at finite and zero temperature for large $q$. In this limit, the $\mathbb{Z}_q$ symmetry of the model becomes a continuous $U(1)$ symmetry. The free energy density and the properties of the phase transition can be obtained from the spectrum of the single-site Hamiltonian $\hat H_s$, which now describes the dynamics of a continuous rotor. In order to take the continuum limit of the clock variable we replace
\begin{equation}
    \sigma\rightarrow e^{i\alpha} \hspace{2cm} \tau \rightarrow e^{-\frac{2\pi}{q}\partial_\alpha}
\end{equation}
such that $\tau$ acts on $\alpha$ as a translation of $2\pi/q$. With this substitution and by expanding $\tau$ to second order in the small parameter $2\pi/q$, we get

\begin{equation}
 H_{s}= 4\pi^2h\left(i\frac{\partial}{\partial \alpha}+\chi\right)^2-2|\lambda| \cos (\alpha +\theta)-2hq^2
\end{equation}
 where $\chi=q \varphi/(2\pi)$.
 Let us distinguish the cases $p=1$ and $p>1$. In the case $p=1$, we have shown that, for any $q>3$ the model has a second order phase transition at the critical value $h_c$ in Eq.~\ref{eq:hc}. Taking the limit $q\rightarrow \infty$ of this expression we find
 
 \begin{equation}
     h_c = \frac{1}{2\pi^2}\frac{1}{1-4\chi^2}.
 \end{equation}
 On the other hand, for $p>1$ the transition is first order and we use approximate methods for locating the transition point: we approximate the potential $-2|\lambda| \cos (\alpha +\theta)$ with a harmonic potential around $\alpha=-\theta$ and find the spectrum to be
\begin{equation}
\label{eq:harmonic}
 E_n=-2|\lambda| -2hq^2+(4
n
+ 2)\pi\sqrt{h|\lambda| }.
\end{equation}

Using Eq.~\ref{eq:f}, we arrive at the following expression for the free energy density
\begin{equation}
    f = (2p-1)\left(\frac{|\lambda| }{p}\right)^{\frac{2p}{2p-1}} - 2h q^2 - 2|\lambda| +2\pi q\sqrt{|\lambda|  h} +\\
    - \frac{1}{\beta}\ln\left(\frac{\sinh(2\pi\beta\sqrt{|\lambda|  h} q)}{\sinh(2\pi \beta\sqrt{|\lambda| h})}\right)
\end{equation}

In the zero-temperature limit ($\beta\to\infty$), it follows from the principle of exponential dominance that the free energy density is given by $ f=(2p-1)|m|^{2p}+E_0 $ for $\chi\neq 1/2$. In order to compute the magnetization ($m_*$) and field ($h_*$) at the transition,
we need to solve two equations simultaneously. The first one is obtained by requiring the free energy of the paramagnet to be equal to the one of the ferromagnet at the transition point,{\it i.e.} $f(\lambda_*, \beta=\infty, h_*) = f(0, \beta=\infty, h_*) $. The second one is arrived at by minimizing the free energy with respect $\lambda$. The result is

 \begin{equation}\label{eq:largeq_critical}
    h_*= \frac{2}{\pi^2}\frac{(2p)^{2p}}{(2p+1)^{2p+1}}   
\hspace{2cm}
     m_*=\frac{2p}{2p+1}
 \end{equation}
 
\begin{figure}
\centering
\includegraphics[width=\textwidth]{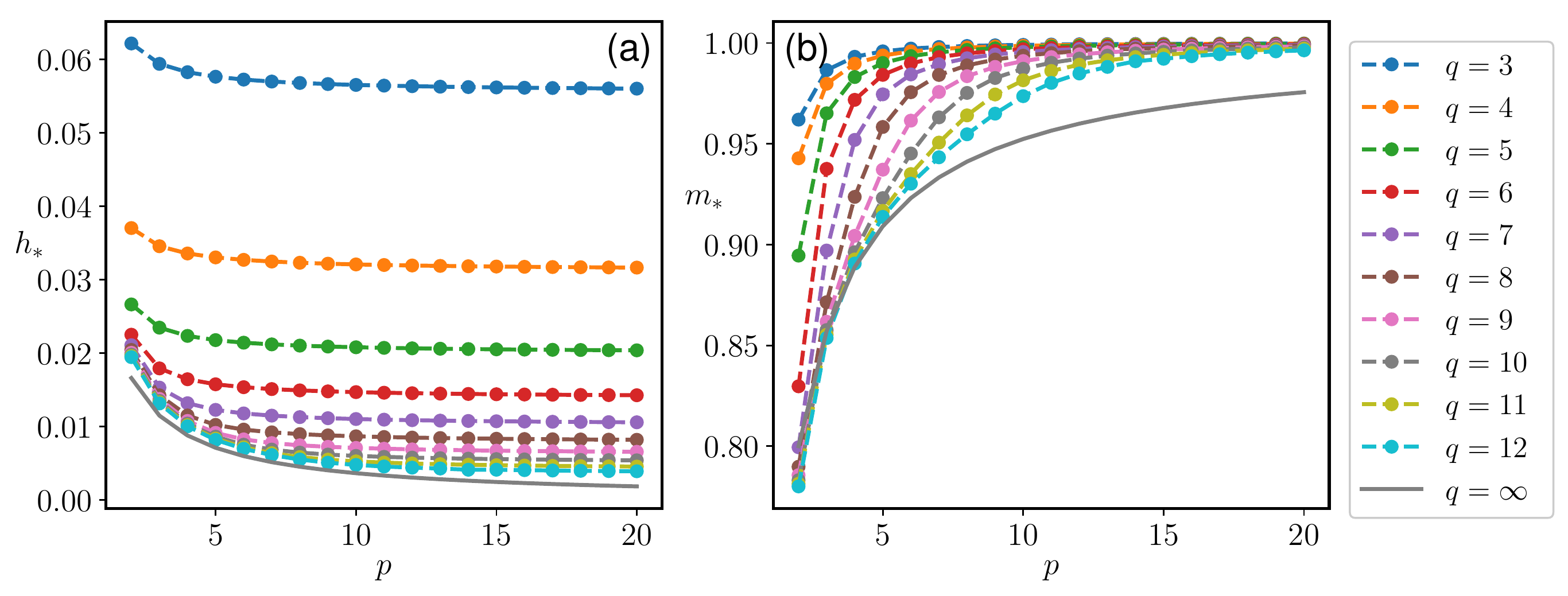}
\caption{Values of the field $h_*$ (panel (a)) and of the magnetization $m_*$ (panel (b)) at the discontinuous phase transition for $p\ge 2$, and different values of $q$. The continuous line is the analytic result obtained in the large $q$ limit (Eq.~\ref{eq:largeq_critical}).}\label{fig:cp}
\end{figure}

By requiring that the transition point belongs to the regime where the harmonic approximation is satisfied (i.e. $\sqrt{h|\lambda|} \ll |\lambda|$), we see that this result is valid when
 $p\gg 1$. In Fig.~\ref{fig:cp} we plot the numerical results for finite $q$ and we see that, as expected, when we increase $q$ they better approximate the analytical results in Eq.~\ref{eq:largeq_critical}.
 
Similarly, the spinoidal field $ h_s $ and magnetization $ m_s $ are computed by solving two equations; the first obtained by requiring that $\frac{\partial^2 f}{\partial \lambda^2} = 0$  and second by minimizing the free energy density with respect to $ \lambda $. We find that
\begin{equation}\label{eq:largeq_spinodal}
  h_{s} = \frac{32 p^2}{\pi^2 } \frac{(2p -1)^{2p - 1}}{(6p-1)^{2 p +1}} 
\hspace{2cm}
    m_{s} = \frac{2p -1 }{6p-1}.
\end{equation}

\section{Conclusions and perspectives}\label{conca:sec}

We have examined the quantum and thermal properties of $\mathbb{Z}_q$-symmetric fully connected clock models,
 and classified the order of their phase transitions.
We showed that the model can have first or second order phase transitions, which depend on 
the (i) chirality $\varphi$ of the model, (ii) order $p$ of the interactions, and (iii) dimensionality $q$
 of the clock variables.
 The full connectivity of the interactions has allowed
us to solve the problem analytically by a combination of mean-field approach
 with perturbation theory up to fourth order.

In our analysis, we have first derived the free energy of the system in a mean-field level, 
 which tends to be exact in the thermodynamic limit for fully connected models.
 In this way we have provided general considerations regarding the possible phase transitions that can occur in the model.
We have applied a Landau-theory argument in the following way. We have expanded the free energy density in terms of 
 the effective longitudinal field $\lambda$ and examined it on the light of the symmetries of the model. In this way we have determined the possible phase transitions the model can have and their respective orders.
The argument relies intimately on the condition that the free energy is an analytic function
 for small effective longitudinal fields, $\lambda \sim 0$.
 We have found that for $p>1$ the possible phase transitions
can only be of first order, while continuous transitions could in principle occur for $p=1$.

 The analyticity condition is satisfied almost always: In the case of a non-analytic free energy around $\lambda \sim 0$ the previous arguments do not apply.
 This is the case of the model at zero temperature with the specific chirality $\varphi = \pi/q$ and $q>2$.
 In this case a different scenario appears and the model has no phase transition to a paramagnetic phase,
 independently of the value of $p$. Remarkably the magnetization remains finite for arbitrarily large fields $h$.

Using perturbation theory up to fourth order, we determined the coefficients 
of the free energy series expansion. This allowed us to perform a quantitative study of the phase transitions and
 delineate the phase diagram of the model for its different parameters $p$, $q$, $\varphi$ and $\beta$.

In the limit $ q \rightarrow \infty$ the $\mathbb{Z}_q$ symmetry of the model becomes a continuous $U(1)$ symmetry. 
In this case we were able to go beyond perturbation theory results, and obtained analytically the free
 energy density of the model with its corresponding critical fields $h_*$ and magnetization $m_*$ (Eq.\eqref{eq:largeq_critical}), as well
 as its and spinodal fields $h_s$ and magnetization $m_s$ 
 (Eq.\eqref{eq:largeq_spinodal}).
 
It is worth mentioning that our results are in agreement with previous works~\cite{federica}, where the case $p=1$, $q=3,4$, $\varphi=0$ was studied numerically.
We remark that the phase structure in the case of infinite-range interactions is much simpler than the one of the one-dimensional short-range model, and has no incommensurate gapless phases.
While in the short-range case for $q=3$, $p=1$ the transition between the trivial phase and the symmetry-breaking phase is second order, here the transition is first order. Also at $\varphi=\pi/3$ what we find is very different from the short-range case, which features a transition from a symmetry-breaking to an incommensurate phase~\cite{zhuang2015}. 
The difference is evident also for larger $q$. For any $q>3$, $p=1$ in the infinite-range model there is a transition from symmetry-breaking to trivial phase but it is second order, in contrast with the already mentioned Kosterlitz-Thouless transition of the one-dimensional self-dual short-range case with $q>4$.
 As a future perspective of this work, it would be interesting to extend the investigation
 to the case of long-range interactions in $d$ dimensions, where it is possible to have
 some spatial dependence of correlations. This case interpolates between one-dimensional short-range and infinite-range cases; studying it would allow to understand the way one moves between two very different phase diagrams. In particular, this step would be important in
order to understand what are the ingredients which allow for the presence of an incommensurate phase,
like the one that arises in short-range interacting clock models in $d = 1$.

\ack{We acknowledge fruitful discussions with A.~Angelone and M.~Dalmonte. F.I. acknowledges the financial support of the Brazilian funding agencies CNPQ (308205/2019-7) and FAPERJ. This work is partly supported by the ERC under
grant number 758329 (AGEnTh).}

\newpage
\section*{Bibliography}
\hspace{0.5cm}
\bibliographystyle{unsrt}
\bibliography{bib.bib}
\newpage
\appendix
\section{Calculation of the pseudo-free energy}\label{sec:trotter}
Here we briefly outline the procedure for computing the Pseudo-free energy of our class of models. The partition function is given by
\begin{equation}
Z(\beta,h)= \text{Tr}[e^{-\beta H}]
\end{equation}

\begin{equation}
Z(\beta,h) = \sum_{\vec{\sigma}}\bra{\vec{\sigma}} e^{\beta N(\hat{m}_{\sigma}\hat{m}_{\sigma}^{\dagger} )^{p} + \beta h q^2N(e^{i \varphi}\hat{m}_{\tau} + e^{-i \varphi}\hat{m}_{\tau}^{\dagger}) }\ket{\vec{\sigma}}
\end{equation}

where $\vec{\sigma} = (\sigma_{1},...,\sigma_{n})$. The Suzuki-Trotter formula is used in order to map the system  onto a classical model with an additional dimension $\alpha$:

\begin{equation}
Z(\beta,h) = \lim _{N_{s} \rightarrow \infty  } \sum_{\vec{\sigma}}\braket{\vec{\sigma} |\big[e^{\beta N(\hat{m}_{\sigma}\hat{m}_{\sigma}^{\dagger})^{p}/N_{s} }e^{\beta N hq^2(e^{i \varphi}\hat{m}_{\tau} + e^{-i \varphi}\hat{m}_{\tau}^{\dagger})/N_{s} }\big]^{N_{s}}|\vec{\sigma}}
\end{equation}
\newline
We introduce $N_s$ closure relations $\mathds{1}(\alpha) = \sum_{\vec{\sigma}(\alpha)} \ket{\vec{\sigma}(\alpha)}\bra{\vec{\sigma}(\alpha)}$ \newline
Where $\alpha$ indicates where the identity is sandwiched. 

\begin{multline}
Z(\beta,h) =\lim _{N_{s} \rightarrow \infty  }\sum_{\vec \sigma(1)...\vec\sigma(N_{s})}\prod_{\alpha=1}^{N_s}
\exp\left\{ \frac{\beta N}{N_s}\left[\left(\sum_{i=1}^N \frac{\sigma_{i}(\alpha)}{N}\right) \,  \left(\sum_{i=1}^{N}\frac{\sigma_i(\alpha)^*}{N}\right)\right]^{p}\right\}\\
\times
\bra{\vec{\sigma}(\alpha)} e^{\beta Nhq^2[e^{i \varphi}\hat{m}_{\tau} + e^{-i \varphi}\hat{m}_{\tau}^{\dagger}]/N_s }\ket{\vec{\sigma}(\alpha +1)}
\end{multline}

where $\sigma(N_s + 1) = \sigma(1)$.
We apply $ N_s $ times the integral representation of the delta function \newline $\int \int \delta(N m_{r}(\alpha) - \Re[\sum _{i}\sigma_{i}])\delta(N m_{im}(\alpha) - \Im[\sum _{i}\sigma_{i}]) f(m_r,m_{im})dm_{im}dm_{r} = f(\Re[\sum \frac{\sigma_{i}(\alpha)}{N}],\Im[\sum \frac{\sigma_{i}(\alpha)}{N}])$

where

\begin{equation}
\delta(N m_{im}(\alpha) - \Im[\sum _{i}\sigma_{i}]) = \int_{-i\infty}^{i\infty} \frac{ d\lambda_{im}}{ \pi i N_{s}/(\beta N)}e^{-\frac{\beta}{N_{s}}2\lambda_{im}(\alpha)(N m_{im}(\alpha) - \Im[\sum _{i}\sigma_{i}(\alpha)])}
\end{equation}

\begin{equation}
\delta(N m_{r}(\alpha) - \Re[\sum _{i}\sigma_{i}]) = \int_{-i\infty}^{i\infty} \frac{ d\lambda_{re}}{ \pi i N_{s}/(\beta N)}e^{-\frac{\beta}{N_{s}}2\lambda_{r}(\alpha)(N m_{r}(\alpha) - \Re[\sum _{i}\sigma_{i}(\alpha)])}
\end{equation}

\begin{align}
Z(\beta,h)=& \lim _{N_{s}\rightarrow \infty  }\int \frac{\prod_{\alpha}dm_{im}(\alpha)dm_{r}(\alpha)d\lambda_{im}(\alpha)d\lambda_{r}(\alpha)}{[ \pi N_{s}/(\beta N)]^2}\\
 & \exp\left[\frac{\beta N}{N_{s}}\left(\sum_{\alpha}\left(m_{r}^{2}(\alpha) + m_{im}^{2}(\alpha)\right)^{p} - 2\lambda_{r}(\alpha)m_{r}(\alpha) -2\lambda_{im}(\alpha)m_{im}(\alpha)\right) \right] \\
 &  \sum_{\vec\sigma(1)...\vec\sigma({N_{s}})}\prod_{\alpha} \bra{\vec{\sigma}(\alpha)} e^{(\beta/N_{s})\sum_{i}\left(hq^2(e^{i\varphi}\tau_{i}+ e^{-i\varphi}\tau_{i}^{\dagger})+ 2\lambda_{im}(\alpha) \Im[\sigma_{i}(\alpha)] + 2\lambda_{Re}(\alpha) \Re[\sigma_{i}(\alpha) ]\right)}\ket{\vec{\sigma}(\alpha +1)} \\
\end{align}
where $ 2\lambda(\alpha)$ is the conjugate variable of the delta function. Using the fact that
\begin{equation}
Tr[A \otimes A\otimes ...\otimes A]  =(Tr[A])^{n} 
\end{equation}
we transform the trace over all spins into that of a single-site problem

\begin{align}
Z(\beta,h)=& \lim _{N_{s}\rightarrow \infty  }\int \frac{\prod_{\alpha}dm_{im}(\alpha)dm_{r}(\alpha)d\lambda_{im}(\alpha)d\lambda_{r}(\alpha)}{[\pi N_{s}/(\beta N)]^2} \\
&\exp\left[\frac{\beta N}{N_{s}}\left(\sum_{\alpha}\left(m_{r}^{2}(\alpha) + m_{im}^{2}(\alpha)\right)^{p} - 2\lambda_{r}(\alpha)m_{r}(\alpha) -2\lambda_{im}(\alpha)m_{im}(\alpha)\right) \right. \\
 &\left.+ \ln\left( \text{Tr}\prod_{\alpha} e^{\beta\left[hq^2(e^{i\varphi}\tau +e^{-i\varphi} \tau^{\dagger}) -i\lambda_{im}(\alpha) (\sigma-\sigma^\dagger) + \lambda_{Re}(\alpha) (\sigma+\sigma^\dagger)\right]/N_s}\right)^N\right]
\end{align}

Here we proceed with the static approximation by setting all the alphas to be equal.

\begin{align}
Z(\beta,h)=& \lim _{N_{s}\rightarrow \infty  }\int \frac{dm_{im}dm_{r}d\lambda_{im}d\lambda_{r}}{[\pi N_{s}/(\beta N)]^2} \\
&\exp\left[\beta N\left(\left(m_{r}^{2} + m_{im}^{2}\right)^{p} - 2\lambda_{r}m_{r} -2\lambda_{im}m_{im}\right) \right. \\
 &\left.+ \ln\left( \text{Tr}\; e^{\beta\left[hq^2(e^{i\varphi}\tau +e^{-i\varphi} \tau^{\dagger}) -i\lambda_{im} (\sigma-\sigma^\dagger) + \lambda_{Re} (\sigma+\sigma^\dagger)\right]}\right)^N\right]
\end{align}

The free energy density is given by the following formula
\begin{equation}
f(\beta,h) = -\frac{1}{\beta N}\ln Z 
\end{equation}

and using the saddle point approximation our integral becomes
\begin{equation} \label{eq:free_energy}
f(\beta,h) = \inf_{m_{im},m_{r}} \text{ext}_{\lambda_{im},\lambda_{r}}[-(m_{r}^2 + m_{im}^2)^{p} + \lambda_{im} m_{im} +\lambda_{r} m_{r} -\frac{1}{\beta}f_{s}(\beta,h,\lambda_{im},\lambda_{re},q))]
\end{equation}

where

\begin{equation}
f_{s}(\beta,h,\lambda_{im},\lambda_{re},q)) = \ln \text{Tr}\left( e^{\beta\left[hq^2(e^{i\varphi}\tau +e^{-i\varphi} \tau^{\dagger}) -i\lambda_{im} (\sigma-\sigma^\dagger) + \lambda_{Re} (\sigma+\sigma^\dagger)\right]}\right)
\end{equation}
Taking partial derivatives of the $f$ with respect to $\lambda_{im},\lambda_{re},m_{im},m_{re} $ and setting the derivatives to be zero, we arrive at the following equations.

\begin{eqnarray}\label{eq:l_of_m}
\lambda_{im} & = &  m_{im} p(m_{im}^2 + m_{re}^2)^{p-1}\\
\lambda_{re} & = &  m_{re} p(m_{im}^2 + m_{re}^2)^{p-1}\\
m_{im}& = & \frac{\partial f_{s}(\beta,h,\lambda_{im},\lambda_{re},q)}{2\beta \partial \lambda_{im}} \\
m_{re}& = & \frac{\partial f_{s}(\beta,h,\lambda_{im},\lambda_{re},q)}{2\beta \partial \lambda_{re}}
\end{eqnarray}

which can be written as

\begin{eqnarray}
\vec{\lambda} & = & \vec{\nabla}_{m} |m|^{2p}\\
\vec{m}& = & \frac{1}{\beta}\vec{\nabla}_{\lambda} g(\beta,h,\lambda_{im},\lambda_{re},q) \\
\end{eqnarray}

moving to radial coordinates we arrive at the following equations

let 
$\vec{\lambda}=(|\lambda|,\phi)$ and $\vec{m}=(|m|,\theta)$

\begin{eqnarray}
|\lambda| & = & 2p|m|^{2p -1 } \\
\phi & = &\theta \\
\theta & = & \frac{\partial g(\beta,h,\lambda,\phi,q)}{2\lambda \beta \partial \phi} \\
|m|& = & \frac{\partial g(\beta,h,|\lambda|,q)}{2\beta \partial |\lambda|}
\end{eqnarray}

\section{Perturbation theory}
\label{app:pertth}
In section \ref{sec:exp} we showed that exploiting the symmetries of the model we can establish which terms can appear in the series expansion of the free energy density. In order to quantitatively determine the coefficients of the series expansion we resort to perturbative calculations. By defining
 \begin{equation}
     \hat H_0 = -hq^2(\hat \tau e^{i\varphi}+ \hat \tau^\dagger e^{-i\varphi}),
     \hspace{2cm}
     \hat V = -(\lambda^* \hat \sigma + \lambda \hat \sigma^\dagger),
 \end{equation}
the free energy density in Eq.~(\ref{eq:f}) can be expressed as $f_s=\sum_{n=0}^\infty f_n$, with

\begin{equation}
  f_{0}=-\frac{1}{\beta}\log \text{Tr } e^{-\beta \hat H_0}, \hspace{1cm}
  f_n = -\frac{1}{\beta} \frac{(-1)^n}{n!}\int_0^\beta dt_1\dots \int_0^\beta dt_n \braket{T [\hat V(t_1)\dots \hat V(t_n)]}_{0,c} \qquad n\ge 1
\end{equation}
where $\hat V(t)=e^{t\hat H_0} \hat V e^{-t\hat H_0}$, and $\braket{T [\hat V(t_1)\dots \hat V(t_n)]}_{0,c}$ is the (imaginary-)time-ordered connected correlation function computed with respect to the unperturbed Hamiltonian $\hat H_0$.

The expansion up to fourth order in the perturbation yields
\begin{equation}
f_{s}=a_0+a_2 \lambda\lambda^*+\delta_{q,2}c_2 \left(\lambda^2+(\lambda^*)^2\right)+\delta_{q,3}c_3 \left(\lambda^3+(\lambda^*)^3\right)+a_4 \lambda^2(\lambda^*)^2+\delta_{q,4}c_4 \left(\lambda^4+(\lambda^*)^4\right)+O(|\lambda|^5).
\end{equation}
We now want to find an expression for the coefficients $a_n$, $c_n$ in terms of the unperturbed eigenvalues $\epsilon_i=-2hq^2\cos(2\pi i/q+\varphi)$ and of $Z_0=\sum_{i=0}^{q-1} e^{-\beta \epsilon_i}$.

Our goal is computing correlation functions of the following form

\begin{equation}
     \int_0^\beta dt_1\dots \int_0^\beta dt_n \braket{T [\hat V(t_1)\dots \hat V(t_n)]}_{0}
     =n!\int_0^\beta dt_n \int_0^{t_n}dt_{n-1}\dots \int_0^{t_{2}} dt_1 \braket{\hat V(t_n)\dots \hat V(t_1)}_{0}.
\end{equation}

We denote by $\ket{i}$ and $\epsilon_i$ with $i=0,\dots q-1$ respectively the eigenstates of $H_0$ and the corresponding eigenvalues. Correlation functions can be computed inserting resolution of the identity $\sum_{i=0}^{q-1}\ket{i}\bra{i}$ as follows
\begin{equation}
    \braket{\hat V(t_n)\dots \hat V(t_1)}_0=\sum_{i_1=0}^{q-1}\dots \sum_{i_n=0}^{q-1}
    e^{-\beta \epsilon_{i_n}} e^{t_n\epsilon_{i_n}}V_{i_ni_{n-1}}e^{-(t_{n}-t_{n-1})\epsilon_{i_{n-1}}}\dots e^{-(t_2-t_{1})\epsilon_{i_1}}V_{i_1i_n}e^{-t_1\epsilon_{i_n}}
\end{equation}
where $V_{ij}=\braket{i|\hat V|j}$
Define $s_n=t_1-t_n+\beta$ and $s_j=t_{j+1}-t_{j}$ for $j=1,\dots n-1$. 

\begin{multline}
\label{eq:diag}
    \int_0^\beta dt_n \int_0^{t_n}dt_{n-1}\dots \int_0^{t_{2}} dt_1 \braket{\hat V(t_n)\dots \hat V(t_1)}_{0}=\\
    =\frac{\beta}{nZ_0} \sum_{i_1=0}^{q-1}\dots \sum_{i_n=0}^{q-1} \int_0^\beta \prod_{j=1}^n ds_j\; \delta\left(\sum_{j=1}^{n}s_j-\beta\right)V_{i_ni_{n-1}}\dots V_{i_1, i_n}\prod_{j=1}^{n}e^{-s_j \epsilon_{i_j}}
\end{multline}

\begin{figure}
\begin{center}
  \includegraphics[width=12cm]{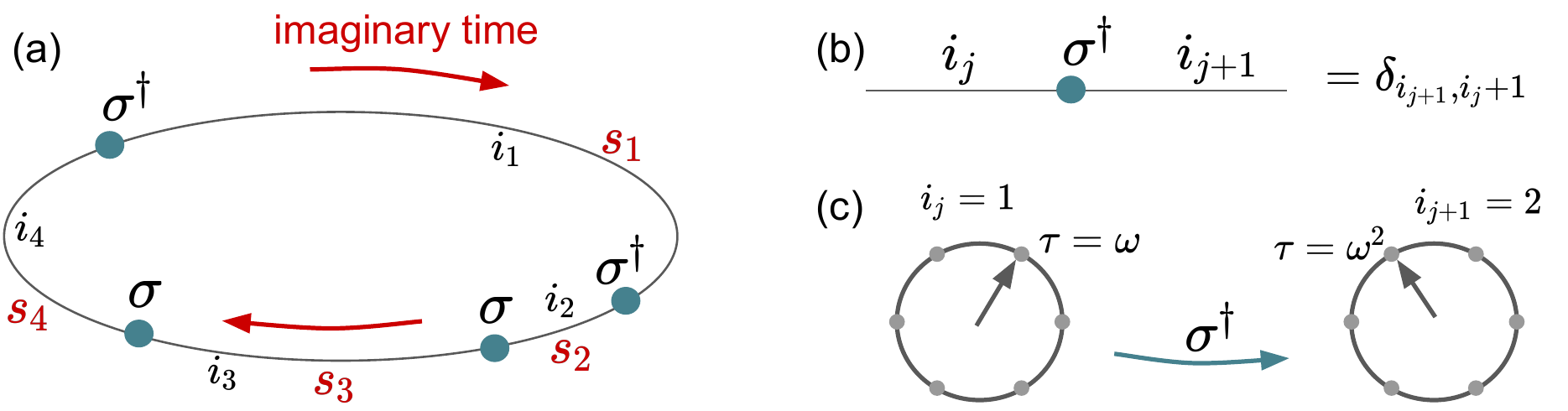}
\end{center}
\caption{(a) Diagrammatic representation of the correlator $\braket{\hat \sigma^\dagger(t_1)\sigma(t_2)\sigma(t_3)\sigma^\dagger(t_4)}_0$ computed as in Eq.~(\ref{eq:diag}). The loop in imaginary time has total length $\beta=s_1+s_2+s_3+s_4$. The indices $i_1,\dots, i_4$ represent the virtual states over which the summation is performed. The green dots are the matrix elements. (b) The matrix elements of $\hat \sigma^\dagger$ in the $\tau$ representation are $\delta$ functions. (c) An example of two states $i_j$, $i_{j+1}$ satisfying the condition $\braket{i_{j+1}|\hat \sigma^\dagger|i_j}=1$.} \label{fig:diagrammatic}
\end{figure}

Since $\hat V=\lambda \hat \sigma+\lambda^*\sigma^{\dagger}$, the last equation can be expressed as a sum of $n$-point correlators of the operators $\hat \sigma$ and $\sigma^\dagger$. Each correlator can be represented with a diagram as in Figure \ref{fig:diagrammatic}: each line represents the imaginary time evolution $e^{-s_j\epsilon_{i_j}}$ and the vertices $\sigma$ and $\sigma^{\dagger}$ flip the state $i_j$ respectively to $i_j-1$ and $i_j+1$. Since the product couples also $i_1$ and $i_n$, it is easy to see that the only correlators which contribute are the ones where, at the end of the cycle, the state comes back to the original one. They correspond to the correlators with an equal number of $\sigma$ and $\sigma^\dagger$, or the ones for which the difference between the numbers of $\sigma$ and $\sigma^\dagger$ operators is a multiple of $q$.

For each diagram we have to compute an integral of the form
\begin{equation}
    I(\beta; \{E_j\})=\int_0^\beta \prod_{j=1}^n ds_j\; \delta\left(\sum_{j=1}^{n}s_j-\beta\right)\prod_{j=1}^{n}e^{-s_j E_j}
\end{equation}
We Laplace transform and obtain
\begin{equation}
    F(\kappa)=\int_0^{+\infty} d \beta\; e^{-\kappa \beta} I(\beta; \{E_j\})=\int_0^{+\infty} \prod_{j=1}^n ds_j\; \prod_{j=1}^{n}e^{-s_j (\kappa+E_j)}=\prod_{j=1}^{n}(k+E_j)^{-1}
    \end{equation}
in the region of convergence $Re(\kappa)>-E_{min}$. We find $I(\beta;\{E_j\})$ as the inverse Laplace transform
\begin{equation}
    I(\beta;\{E_j\})=\frac{1}{2\pi i} \lim_{T\rightarrow +\infty} \int_{\gamma-iT}^{\gamma+iT}e^{\kappa \beta} \prod_{j=1}^{n}(k+E_j)^{-1}
\end{equation}
with $\gamma>-E_{min}$. The integral is then easily computed using the residue theorem.
\begin{equation}
    I(\beta;\{E_j\})=\sum_{j}\mathrm{Res}(f, -E_j)
\end{equation}
with $f(z)=e^{z \beta} \prod_{j=1}^{n}(z+E_j)^{-1}$ and the sum counts each pole once.

\subsection{Order $n=2$}
For $q>2$, it is easy to see that only two diagrams which contribute, and they give the same integral
\begin{multline}
\int_0^\beta dt_2\int_0^{t_2} dt_1 \braket{\hat V(t_2)\hat V(t_1)}_{0}=
|\lambda|^2\left(\int_0^\beta dt_2\int_0^{t_2} dt_1 \braket{\hat \sigma(t_2)\hat \sigma^\dagger(t_1)}_0 + \int_0^\beta dt_2\int_0^{t_2} dt_1 \braket{\hat \sigma^\dagger(t_2)\hat \sigma(t_1)}_0 \right)
\\=2|\lambda|^2\int_0^\beta dt_2\int_0^{t_2} dt_1 \braket{\hat \sigma(t_2)\hat \sigma^\dagger(t_1)}_0
\label{eq:2corr}
\end{multline}

\begin{equation}
\int_0^\beta dt_2\int_0^{t_2} dt_1 \braket{\hat \sigma(t_2)\hat \sigma^\dagger(t_1)}_0=\frac{\beta}{2Z_0}\sum_{i=0}^{q-1}I(\beta; \epsilon_i, \epsilon_{i+1})=-\frac{\beta}{2Z_0}\sum_{i=0}^{q-1}\frac{e^{-\beta \epsilon_{i+1}}-e^{-\beta \epsilon_i}}{\epsilon_{i+1}-\epsilon_i}
\label{eq:2}
\end{equation}
The last equation is obtained for the case of simple poles. If, for some state $i$, $\epsilon_i=\epsilon_{i+1}$, the equation is valid with the substitution $\frac{e^{-\beta \epsilon_{i+1}}-e^{-\beta \epsilon_i}}{\epsilon_{i+1}-\epsilon_i}\rightarrow -\beta$.

The second order term in the expansion of the free energy is
\begin{equation}
    f_2=a_2|\lambda|^2, \qquad a_2=\frac{1}{Z_0} \sum_{i=0}^{q-1}\frac{e^{-\beta \epsilon_{i+1}}-e^{-\beta \epsilon_i}}{\epsilon_{i+1}-\epsilon_i}<0.
\end{equation}

\subsection{Order $n=3$}
A third order term is present only for $q=3$. In that case we get
\begin{equation}
\int_I d\mathbf{t}
 \braket{\hat V(t_3)\hat V(t_2)\hat V(t_1)}_{0}
=\lambda^3\int_I d\mathbf{t} \braket{\hat \sigma(t_3)\hat \sigma(t_2)\hat \sigma(t_1)}_0+(\lambda^*)^3 \int_I d\mathbf{t} \braket{\hat \sigma^\dagger(t_3)\hat \sigma^\dagger(t_2)\hat \sigma^\dagger(t_1)}_0
\end{equation}
where the notation $\int_I d\mathbf{t}$ is used to denote the integration over the region $0\le t_1\le t_2\le t_3 \le \beta$. We obtain

\begin{equation}
\int_I d\mathbf{t}
\braket{\hat \sigma(t_3)\hat \sigma(t_2)\hat \sigma(t_1)}_0=\frac{\beta}{Z_0} I(\beta; \epsilon_0, \epsilon_1, \epsilon_2),
\end{equation}
\begin{equation}
    I(\beta; \epsilon_0, \epsilon_1, \epsilon_2)=\frac{e^{-\beta \epsilon_0}}{(\epsilon_1-\epsilon_0)(\epsilon_2-\epsilon_0)}+
    \frac{e^{-\beta \epsilon_1}}{(\epsilon_2-\epsilon_1)(\epsilon_0-\epsilon_1)}+\frac{e^{-\beta \epsilon_2}}{(\epsilon_0-\epsilon_2)(\epsilon_1-\epsilon_2)},
\end{equation}
from which we get the term of the free energy density
\begin{equation}
f_3 =c_3\delta_{q,3}(\lambda^3+(\lambda^*)^3),\qquad c_3=\frac{I(\beta; \epsilon_0, \epsilon_1, \epsilon_2)}{Z_0}     
\end{equation}

\subsection{Order $n=4$}
If $q \neq 4$, the only diagrams that contribute are

\begin{multline}
\int_I d\mathbf{t} \braket{\hat V(t_4)\hat V(t_3)\hat V(t_2)\hat V(t_1)}_{0}=\\
=4|\lambda|^4\int_I d\mathbf{t} \braket{\hat \sigma^\dagger(t_4)\hat \sigma^\dagger(t_3)\hat \sigma(t_2)\hat \sigma(t_1)}_0+2|\lambda|^4\int_I d\mathbf{t} \braket{\hat \sigma^\dagger(t_4)\hat \sigma(t_3)\hat \sigma^\dagger(t_2)\hat \sigma(t_1)}_0
\label{eq:4corr}
\end{multline}
(now the integration is over the region $0\le t_1\le t_2\le t_3 \le t_4 \le \beta$).

\begin{equation}
\label{eq:4a}
\int_I d\mathbf{t} \braket{\hat \sigma^\dagger(t_4)\hat \sigma^\dagger(t_3)\hat \sigma(t_2)\hat \sigma(t_1)}_0=\frac{\beta}{4Z_0}\sum_{i=0}^{q-1}I(\beta; \epsilon_{i-1}, \epsilon_i, \epsilon_{i+1}, \epsilon_i)
\end{equation}

\begin{multline}
    I(\beta; \epsilon_{i-1}, \epsilon_i, \epsilon_{i+1}, \epsilon_i)=\frac{\beta e^{-\beta \epsilon_i}}{(\epsilon_{i+1}-\epsilon_i)(\epsilon_{i-1}-\epsilon_i)}-\frac{e^{-\beta\epsilon_i}}{(\epsilon_{i+1}-\epsilon_i)(\epsilon_{i-1}-\epsilon_i)^2}-\frac{e^{-\beta\epsilon_i}}{(\epsilon_{i+1}-\epsilon_i)^2(\epsilon_{i-1}-\epsilon_i)}
    \\
    +\frac{e^{-\beta\epsilon_{i+1}}}{(\epsilon_i-\epsilon_{i+1})^2(\epsilon_{i-1}-\epsilon_{i+1})}
    +\frac{e^{-\beta\epsilon_{i-1}}}{(\epsilon_i-\epsilon_{i-1})^2(\epsilon_{i+1}-\epsilon_{i-1})}
\end{multline}

\begin{equation}
\int_I d\mathbf{t} \braket{\hat \sigma^\dagger(t_4)\hat \sigma(t_3)\hat \sigma^\dagger(t_2)\hat \sigma(t_1)}_0=\frac{\beta}{4Z_0}\sum_{i=0}^{q-1}I(\beta; \epsilon_{i+1}, \epsilon_i, \epsilon_{i+1}, \epsilon_i)
\end{equation}

\begin{equation}
    I(\beta; \epsilon_{i+1}, \epsilon_i, \epsilon_{i+1}, \epsilon_i)=\frac{\beta(e^{-\beta \epsilon_i}+e^{-\beta \epsilon_{i+1}})}{(\epsilon_{i+1}-\epsilon_i)^2}+\frac{2(e^{-\beta \epsilon_{i+1}}-e^{-\beta \epsilon_{i}})}{(\epsilon_{i+1}-\epsilon_i)^3}
\end{equation}

For the case $q=4$, there is an additional term coming from the diagram
\begin{equation}
\label{eq:4b}
    \int_I d\mathbf{t} \braket{\hat \sigma(t_4)\hat \sigma(t_3)\hat \sigma(t_2)\hat \sigma(t_1)}_0=\frac{\beta}{Z_0}I(\beta; \epsilon_0, \epsilon_1, \epsilon_2, \epsilon_3).
\end{equation}

by inserting Eqs. \ref{eq:2corr} and \ref{eq:4corr} in 
\begin{equation}
    f_4=-\frac{1}{4!\beta}\left[4!\int_I d\mathbf{t} \braket{\hat V(t_4)\hat V(t_3)\hat V(t_2)\hat V(t_1)}_{0}- 3\cdot (2!)^2\left(\int_0^\beta dt_2 \int_0^{t_2} dt_1 \braket{\hat V(t_2)\hat V(t_1)}_{0}\right)^2\right]
\end{equation}

and using Eqs. \ref{eq:4a}, \ref{eq:4b} and \ref{eq:2},
we obtain the final result
\begin{equation}
    f_4 =a_4|\lambda|^4+c_4\delta_{q,4}(\lambda^4+(\lambda^*)^4)
\end{equation}
with
\begin{equation}
    a_4 = \frac{1}{2Z_0}\left[-\sum_{i=0}^{q-1} \bigg(2I(\beta; \epsilon_{i-1}, \epsilon_i, \epsilon_{i+1}, \epsilon_i)+I(\beta; \epsilon_{i+1}, \epsilon_i, \epsilon_{i+1}, \epsilon_i)\bigg)+\frac{\beta}{Z_0}\left(\sum_{i=0}^{q-1}I(\beta; \epsilon_i, \epsilon_{i+1})\right)^2\right],
\end{equation}
\begin{equation}
    c_4 = -\frac{I(\beta; \epsilon_0, \epsilon_1, \epsilon_2, \epsilon_3)}{Z_0}.
\end{equation}

\section{Proof of first order transition for $p=1$, $q=3$}
\label{app:firstorder}
We here prove that the transition is of the first order for $p=1$, $q=3$.
Up to fourth order terms the free energy density has the form
\begin{equation}
f(\lambda\equiv|\lambda|e^{i\theta})=C_0+(1+a_2) |\lambda|^2+2 c_3 \cos (3\theta) |\lambda|^3+a_4 |\lambda|^4+O(|\lambda|^6)
\end{equation}
where $a_4$ is positive for $h\simeq h_c$. The transition is first order if for a certain $a_2>-1$ there exists some $\lambda_*$ such that $f(\lambda_*)<f(0)$. Neglecting terms beyond fourth order, we find that this condition reduces to a second degree inequality, which has solutions only if
\begin{equation}
    \Delta = (c_3 \cos(3\theta))^2-a_4 (1+a_2)>0
\end{equation}
For $h\rightarrow h_c^+$ we have $(1+a_2)\rightarrow 0^+$, so $\Delta>0$ holds for $h$ sufficiently close to $h_c$. We fix $\theta=0$ if $c_3<0$ and $\theta=\pi/3$ if $c_3>0$, such that $2c_3\cos (3\theta)=-2|c_3|$. Then, to lowest order in $1+a_2$, the inequality is satisfied for
\begin{equation}
    \frac{1+a_2}{2|c_3|}<|\lambda|<\frac{2|c_3|}{a_4}.\label{eq:ineqsol}
\end{equation}
At this point one might be concerned that this solution may break when including terms beyond the fourth order. The crucial observation is that for $h\rightarrow h_c$, the lower extremum of the range in Eq.~\ref{eq:ineqsol} $(1+a_2)/2|c_3|$ goes to 0: hence we find solutions for arbitrarily small values of $|\lambda|$, where higher order terms are negligible. Therefore, the existence of $\lambda_*$ such that $f(\lambda_*)<f(0)$ can always be satisfied, for some $h>h_c$, and the transition is of the first order.
\end{document}